# STUDY OF THE IMPACT OF CLIMATE CHANGE ON PRECIPITATION IN PARIS AREA USING METHOD BASED ON ITERATIVE MULTISCALE DYNAMIC TIME WARPING (IMs-DTW)

Mohamed Djallel Dilmi[1], Laurent Barthès[1], Cécile Mallet[1], Aymeric Chazottes[1]

*Abstract*— **Studying the impact of climate change on precipitation is constrained by finding a way to evaluate the evolution of precipitation variability over time. Classical approaches (feature-based) have shown their limitations for this issue due to the intermittent and irregular nature of precipitation. In this study, we present a novel variant of the Dynamic time warping method quantifying the dissimilarity between two rainfall time series based on shapes comparisons, for clustering annual time series recorded at daily scale. This shape-based approach considers the whole information (variability, trends and intermittency…). We further labeled each cluster using a feature-based approach. While testing the proposed approach on the time series of Paris-Montsouris, we found that the precipitation variability increased over the years in Paris area.**

## I. MOTIVATION

Climate change is a reality widely recognized today [1], [2]. It corresponds to a lasting change in parameter statistics such as those of the distribution of temperatures or precipitations over a period of several decades to several thousand years. These changes may be due to processes intrinsic to the Earth, external influences such as solar radiation or, more recently, human activities. The anthropogenic climate change responsible for global warming is the result of greenhouse gas emissions generated by human activities, which alter the composition of the planet's atmosphere. Due to the direct correlation between greenhouse gas emissions (particularly CO2) and temperature, the impact of climate change on temperature has been studied extensively. Nevertheless, its impact on precipitation remains unclear; it is sometimes assumed that precipitation variability does not change in a warming climate [3], [4], or that mean precipitation and its variability change at the same rate [5], or even that the precipitation variability increases in warmer climate [6], [7]. It remains difficult to observe and predict any impact of climate change due to the intermittent and irregular nature of precipitation [8], [9]. These different works show the difficulty to conclude about the evolution of rainfall variability in the context of climate change.

Only rain gauges provide continuous observation of precipitation throughout the last century. They provide daily time series of uninterrupted rainfall with valuable information on the history of the regional climate and thus allow to study the evolution of precipitation over time. Since precipitation is an intermittent phenomenon that takes the form of precipitating events, the daily scale, close to the duration of the events, is well suited for the study of precipitation behavior over the past 150 years. In climatology, it is customary to split the time series into hydrological years (from September to August) and compare them when trying to detect an evolution.

Using a feature-based approach (e.g., mean rain rate, maximum values of rain rates …) to characterize the variability of rainfall time series and describe its evolution over years is sub-optimal [10]

Corresponding author: M. D. DILMI djallel.dilmi@latmos.ipsl.fr
[1] LATMOS/CNRS/UVSQ/Université Paris-Saclay, 11 boulevard d'Alembert, 78280 Guyancourt, France

because there is more information (about how the variability evolves) that is available if the entire annual time series is used than if only some extracted features are used. Dilmi et al. [11] proposed a shape-based approach for rainfall time series comparison that considers the whole information (variability, tends and intermittency…) of time series while quantifying the dissimilarity between them, the details of the method are presented in part II. In part III, we applied this approach to cluster annual time series observed at daily scale, which allowed us to track the evolution of precipitation over the years. The interpretation of the raw outputs of this approach is performed by applying a feature-based approach.

In this paper, we describe a multi-step approach that combines the shape-based clustering algorithm and the feature-based approach to overcome the individual limitations of both approaches, in order to better investigate and describe the evolution of precipitation variability over the years.

## II. Method

Given a long time series of daily-scale rain rates RR [mm.h$^{-1}$] that was split into a set of **N** hydrological annual time series, **we first compare all pairs (i,j) of the NxN annual time series.** The comparison method we used is the iterative multiscale dynamic time warping (IMs-DTW) method [11]. Described as a time-normalized distance between two rainfall time series, it is a variant of the Dynamic time warping [12][13] which searches for an optimal match (called alignment) between the two time series. It allows to stretch and compress some sub-sections of the time axis at different time scale while respecting some constraints in order to minimize the dissimilarity between the compared time series. Then, it assigns a dissimilarity score based on the found alignment to the compared pair of annual time series. Finally, all dissimilarity scores are ranked in a matrix called dissimilarity matrix [NxN].

**The second step is using the dissimilarity scores to perform clustering of annual time series.** The clustering method we used is K-medoids [14][15] which is a variant of K-means where the cluster medoid is defined to be the closest annual time series to the set of annual time series in the cluster. The clustering method is as follows:

1) Randomly initialize K medoids $m_k$ $with$ $k = 1 \dots K$ (K annual time series), one for each cluster $C_k$.
2) For every annual time series in the dataset, find the nearest cluster's medoid based on the dissimilarity score calculated above, and then assign the time series to the corresponding cluster.
3) Find the central medoid $m_k$ for new clusters $C_k$ $k = 1 \dots K$ which minimize the formula : $m_k = \arg\min_{j \in C_k}(\sum_{i \in C_k} dissim(i,j)^2)$ with dissim(i,j) representing the dissimilarity score between the two annual time series i an j.
4) Repeating steps 2 and 3 until there are no more changes.

**The third step consists of labeling each cluster** separately based on a detailed analysis of its medoid and the characteristics it shares with the cluster. Extracted precipitation indices were used in this step (ex. Maximum and standard-deviation of rain rates over the year, precipitation amount, precipitation duration, consecutive wet days, consecutive dray days …) [1], as well as other environmental features that can be easily interpreted [1] and historical rain bibliography [7][16].

Finally, after labeling each cluster, **we analyze the evolution of their frequencies over time.** For this we split the time axis into several successive time intervals large enough and from each cluster calculated the frequency of its presence, and sought to identify trends. In general, a period of 30 years is suggested for intervals by the world Meteorological Organization (WOM).

## III. Evaluation

We chose to apply the approach to study the area of Paris, specifically on the station of Paris-Montsouris in France, which provides an unbroken time series of daily rain gauge measurements of precipitation since 1873, validated by the World Meteorological Organization (WMO).

Following the approach proposed above, we carried out an analysis of the time series measured between Sept.1st, 1873 and Aug. 31st, 2019 at the daily scale, split into 146 annual time series.

The comparison of all annual time series by pairs using the IMs-DTW provides a set of alignments. Figure 1 illustrates, by way of example, the alignment obtained between the two annual time series representing years 1890 and 1900. The alignment structure ensures the association of the peaks and the rainy periods of the two years, we note that the associations of the matching present relatively small offsets with up to two weeks. This observation is true for all the alignments proposed by the IMS-DTW between the 146 annual times series.

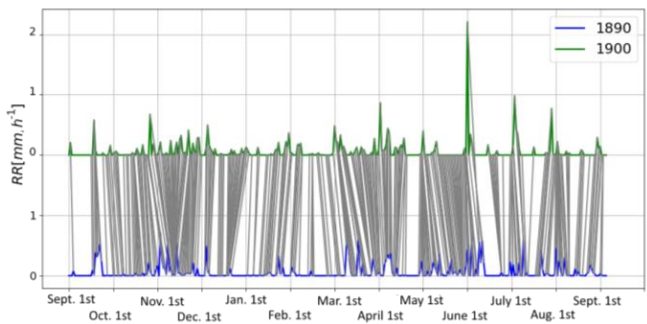

**Figure 1: The alignment that links the two annual rainfall time series representing the two years 1890 and 1900 measured at the station Paris-Montsouris**

The small distortions of time axis (less than 2 weeks) between the annual time series implicitly guarantee the seasonal comparison of different years (summer compared to summer and winter to winter). Sometimes, a season can be shifted by a few weeks, and the IMS-DTW not only allows to correct but also to precisely evaluate this shift.

The dissimilarity scores are ranked in the dissimilarity matrix D[146x146] presented in Figure 2. The higher the value, the more dissumilar two years are.

All observations reported on "climate change in Paris" published by Météo-France [14] are verified and visible on the dissimilarity matrix D. Among the annual time series that are standing out on the dissimilarity matrix, we can mention:

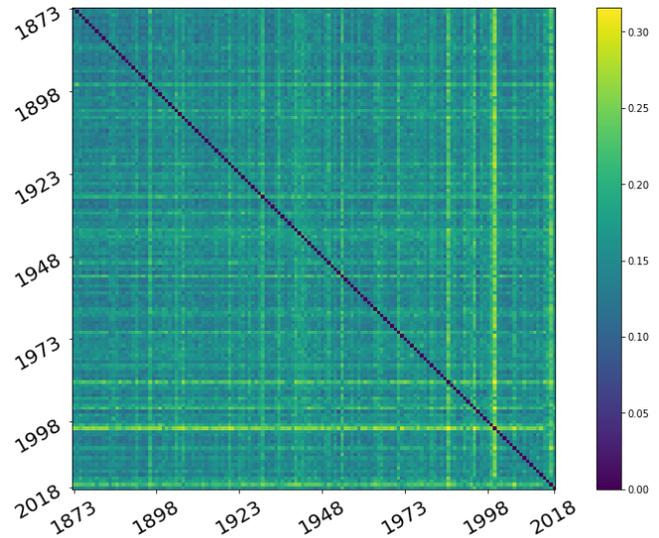

**Figure 2: the dissimilarity matrix D (146x146) between the 146 annual time series measured in Paris-Montsouris between 1873 and 2019**

1) The annual time series 1906 (#34) known in the Parisian history for the flood of the Seine River and listed as an exceptional year, presents important dissimilarities from other years.

2) There is also a band that marks the four years corresponding to the period from September 1st, 1940, to August 31st, 1944 (# 68 to 71). This could be explained by disturbances in data collection during the war period.

3) The annual time series 1954 (# 82), also listed as an exceptional year for the flood of the Seine River.

4) The annual time series 2000 (# 128) has strong dissimilarities (the maximums) compared to the other annual time series of the study.

The time series clustering into four clusters ($K = 4$) of annual time series provides four central medoids. Their analysis, associated with that of the corresponding cluster, allows the following ascertainments:

-        The central medoid of the first cluster (#C1) is 1953 (#81), this cluster groups 34 annual time series. All are characterized by particularly low precipitation intensity, variability and water accumulation; as well as longer periods without rain.

These characteristics correspond to the label **"drought years"**.

- The central medoid of the fourth cluster (C4) is 1884 (#12), this cluster (#C4) groups 28 annual time series, all are characterized by extreme variability and violent thunderstorms in summer that recorded exceptional values for rain rates and variability. These characteristics correspond to the label **"remarkable years with extreme variability"**. All the remarkable years reported in MétéoFrance's report [14] belong to this cluster.

- The two clusters C2 (1995 (#123) as medoid) and C3 (1931 (#59) as medoid) were difficult to labeling because the two of them group normal years.

Considering the appearance frequency of each cluster in seven successive 26 years' time intervals between 1873 and 2019 allows studying the possible evolution of precipitation (see table 1).

| Temporal Window | | Cluster | | | |
|---|---|---|---|---|---|
| | | C1 | C2 | C3 | C4 |
| 1873 | 1899 | 29 | 19 | 16 | 07 |
| 1892 | 1918 | 21 | 19 | 19 | 14 |
| 1911 | 1937 | 18 | 15 | 29 | 11 |
| 1930 | 1956 | 18 | 11 | 24 | 25 |
| 1949 | 1975 | 21 | 23 | 81 | 21 |
| 1968 | 1994 | 18 | 23 | 11 | 21 |
| 1987 | 2018 | 09 | 28 | 16 | 36 |
| Gradient | | ↘ | ↘↗ | ↗↘ | ↗ |

**Table 1. Temporal evolution of frequencies for each cluster [% by column]. The temporal window is 26 years length with an overlap of 7 years.**

With a coarse time window, we see that the presence of the first cluster C1 (**drought years**) decreases over time while that of C4 (remarkable years with extreme variability) increases or stagnates at the same time which corroborates the assumption that the precipitation variability increases over time (# warmer climate) between 1873 and 2019.

## IV. Limitation and Current Work

The proposed method shows that there exists a temporal evolution of precipitation when studying Paris-Montsouris time series. However limitations of the current work include the fact that we have not yet demonstrated the robustness of the results in table 1, i.e. we have not yet tested whether other rainfall time series measured in Paris area could lead to similar results nor demonstrated that results show a real impact of climate change rather than just representing a temporal correlation. We are currently working on both of these topics, and are furthermore exploring the correlation between these clusters and environmental parameters (such as temperature). We also plan to extend the study area and to use other rainfall time series. The proposed unsupervised classification of annual time series based on a new dissimilarity measure seems particularly well adapted to the intermittent nature of precipitation time-series. The K medoid approach was used so as to facilitate the interpretation step, which allowed to label each year in function of theire structure of annual precipitation. Finally, the analysis of the frequency of these labels of years from 1873 to the current day makes it possible to describe the annual evolution of precipitation observed at a daily scale.